\documentclass{article}
\usepackage{spconf,amsmath,amssymb,amsfonts,graphicx}
\usepackage{cite}
\usepackage{textcomp}
\usepackage{xcolor}
\usepackage{url}
\usepackage{makecell}
\usepackage{tikz}
\usetikzlibrary{arrows.meta, positioning, calc}

\newif\ifblindreview
\blindreviewfalse

\title{Vaani Benchmark V1.0: An Inclusive Multimodal Benchmark Dataset for Hindi}

\ifblindreview
\name{Anonymous Submission}
\address{Affiliations omitted for double-blind review}
\else
\makeatletter
\gdef\@name{%
  {\em Sujith Pulikodan\textsuperscript{1},
       Agneedh Basu\textsuperscript{1},
       Saurabh Kumar\textsuperscript{2},
       Pranav Bhat\textsuperscript{1}}\\
  {\em Pavan Kumar J\textsuperscript{1},
       Nihar Desai\textsuperscript{1},
       Prasanta K. Ghosh\textsuperscript{2}}}
\gdef\@address{%
  \textsuperscript{1}\emph{AI \& Robotics Technology Park (ARTPARK), I-Hub @ IISc}, Bangalore, India\\
  \textsuperscript{2}\emph{Department of Electrical Engineering, Indian Institute of Science}, Bangalore, India\\
  \texttt{sujith@artpark.in}, \texttt{spirelab.ee@iisc.ac.in}}
\makeatother
\fi

\begin{document}
\ninept
\maketitle
\begin{abstract}
Benchmarking is critical for the systematic evaluation of machine learning systems. While several open-source datasets are available for Hindi, existing benchmarks remain limited in terms of modality, geographic diversity, demographic representation, and transcription robustness. We introduce an inclusive, multimodal Hindi benchmark dataset collected from 102 districts across India. The dataset consists of spontaneous speech elicited using image prompts and recorded under real-world acoustic conditions across diverse demographic groups. Each audio segment is associated with the image prompt that elicited it and is annotated with three independent transcriptions, enabling multi-reference evaluation that accounts for permissible orthographic and lexical variations. We show that single-reference evaluation overstates errors, while
multi-reference evaluation provides a more robust, inclusive, and
realistic assessment of automatic speech recognition (ASR) systems. The pairing of each utterance with its eliciting image further enables image retrieval evaluation using both speech and text queries. We evaluate multiple multimodal embedding models for image retrieval and report their performance on the combined speech–image and text–image retrieval tasks. The results reveal a  performance gap between text-based and audio-based retrieval, with text-based retrieval consistently achieving higher performance.

\end{abstract}
\section{Introduction}
\label{sec:intro}

Benchmarking is critical for evaluating model performance under unseen conditions, comparing systems, and assessing their real-world applicability. State-of-the-art ASR systems remain sensitive to variations in gender, age, accent, and other speaker characteristics \cite{feng2021quantifyingbiasautomaticspeech,ngueajio2022hey}. This challenge is particularly important for Hindi, which, according to the 2011 Census of India, is the most widely spoken language in the country, with 43.63\% of the population reporting Hindi or its dialects as their first language \cite{census2011india}. Hindi spans diverse regions and dialects, forming a broad linguistic continuum that requires benchmarks with adequate geographic, demographic, and linguistic coverage. Similarly, image and speech retrieval systems require evaluation on datasets that reflect local linguistic, cultural, and visual contexts. However, existing benchmarks often provide limited coverage of such diversity, motivating the development of comprehensive and locally relevant multimodal benchmarks for Hindi.

\begin{figure*}[t]
\centering
\resizebox{\textwidth}{!}{%
\begin{tikzpicture}[
  base/.style   = {draw, rounded corners=3pt, align=center, font=\footnotesize,
                   text width=2.05cm, minimum height=0.8cm, inner sep=2pt,
                   execute at begin node={\hyphenpenalty=10000\exhyphenpenalty=10000
                                          \hbadness=10000}},
  bluebox/.style   = {base, fill=blue!12,   draw=blue!55},
  yellowbox/.style = {base, fill=yellow!25, draw=orange!70},
  greenbox/.style  = {base, fill=green!18,  draw=green!45!black},
  redbox/.style    = {base, fill=red!13,    draw=red!55},
  purplebox/.style = {base, fill=violet!13, draw=violet!55},
  arr/.style    = {-{Stealth[length=2mm]}, semithick, draw=black!80},
  line/.style   = {semithick, draw=black!80}
]

\node[bluebox, text width=3.05cm] (land)  at (0,0)        {Identify Locally Relevant\\Landmarks / Topics};
\node[bluebox]                    (img)   at (3.30,0)     {Image Collection};
\node[greenbox]                   (iauto) at (3.30,-1.75) {Automated Check};
\node[redbox]                     (iman)  at (0,-1.75)    {Manual Check};

\node[bluebox]                    (spch)  at (0,-3.50)    {Spontaneous\\Speech Collection};
\node[yellowbox, text width=2.4cm](aq)    at (3.30,-3.50) {Audio Quality \&\\Format Check};

\foreach \i/\y in {1/0, 2/-1.75, 3/-3.50}{
  \node[purplebox] (t\i)    at (6.30,\y)  {Transcription \i};
  \node[greenbox]  (qc\i a) at (9.00,\y)  {Automated QC};
  \node[redbox]    (mc\i)   at (11.70,\y) {Independent\\Manual Correction};
  \node[greenbox]  (qc\i b) at (14.40,\y) {Automated QC};
}

\node[greenbox]                    (score)   at (17.10,-1.75) {Inconsistency\\Scoring};
\node[yellowbox]                   (recheck) at (18.95,-3.50) {Manual Recheck\\+ Automated QC};
\node[bluebox, text width=2.6cm]   (final)   at (21.70,-1.75) {Final Output\\3 Transcriptions\\per Segment};

\draw[arr] (land)  -- (img);
\draw[arr] (img)   -- (iauto);
\draw[arr] (iauto) -- (iman);
\draw[arr] (iman)  -- node[right, font=\scriptsize, pos=0.42] {Images} (spch);
\draw[arr] (spch)  -- (aq);

\coordinate (br) at (4.90,-3.50);
\draw[line] (aq.east) -- (br);
\draw[arr]  (br) |- (t1.west);
\draw[arr]  (br) |- (t2.west);
\draw[arr]  (br) -- (t3.west);

\foreach \i in {1,2,3}{
  \draw[arr] (t\i)    -- (qc\i a);
  \draw[arr] (qc\i a) -- (mc\i);
  \draw[arr] (mc\i)   -- (qc\i b);
}

\coordinate (bus) at (15.75,-1.75);
\draw[line] (qc1b.east) -| (bus);
\draw[line] (qc3b.east) -| (bus);
\draw[arr]  (qc2b.east) -- (score.west);

\draw[arr] (score) -- (final);
\draw[arr] (score.south) |- (recheck.west);
\draw[arr] (recheck.east) -| (final.south);

\end{tikzpicture}
}
\caption{Benchmark Data Preparation Process}
\label{fig:pipeline}
\end{figure*}

Several benchmark datasets have been released to evaluate Hindi ASR either as part of larger corpora or through independent initiatives, such as IndicSUPERB \cite{javed2023indicsuperb}, LAHAJA \cite{javed2024lahaja} and Vistaar \cite{bhogale2023vistaar}. LAHAJA comprises both read and extemporaneous speech across a diverse range of topics and use cases, totaling 12.5 hours of Hindi audio collected from 132 speakers spanning 83 districts of India. In addition, multiple data collection efforts---including FLEURS \cite{conneau2023fleurs}, MUCS \cite{Diwan_2021} and GramVaani \cite{kumar2022gram} ---have released evaluation sets that can be used to benchmark ASR performance. Despite the availability of numerous datasets, significant gaps remain in open benchmark datasets, particularly in terms of comprehensive regional and dialectal coverage.

The widely used ASR metric, Word Error Rate (WER), is sensitive to variations in valid reference transcriptions, making single-reference evaluation potentially unfair \cite{mcnamara2024styleagnosticevaluationasrusing,ali2015multi}. This is particularly relevant for Hindi, where existing benchmarks often provide only one reference and fail to capture natural transcriptional variability. Similarly, code-switching is prevalent in Indic languages, yet current benchmarks often lack adequate code-switched speech or use inconsistent script conventions, limiting fair evaluation \cite{barnali2017code,agro2025code}. ASR performance is also strongly affected by background noise, but many benchmarks lack explicit noise-event annotations needed to assess robustness under realistic acoustic conditions \cite{gong1995speech}. These limitations highlight the need for benchmarks that incorporate multiple human references, code-switching, and diverse noise conditions to enable more reliable and comprehensive ASR evaluation. Hindi image--text \cite{sharma2026chitravivran} and speech--image datasets exist \cite{harwath2018vision}, but resources that jointly provide images, Hindi text, and naturally spoken Hindi audio remain limited. Recent multimodal models increasingly integrate multiple modalities \cite{comanici2025gemini}, highlighting the need for well-aligned datasets for systematic evaluation. 

To address these gaps, we introduce a multimodal benchmark for evaluating Hindi ASR and text and speech-based image retrieval. The benchmark contains paired speech--image data with three independent reference transcriptions per audio segment, enabling multi-reference ASR evaluation. Data were collected from 102 districts across 20 Indian states and 2 Union Territories, providing broad geographic and demographic coverage and capturing regional variation in Hindi speech.

\section{Benchmark Dataset}
\label{sec:dataset}

The data were collected as part of the Vaani Project \cite{vaani2025} and comprise spontaneous speech elicited by asking speakers to describe images. The image set includes both district-specific images and images shared across multiple districts. Carefully designed guidelines were followed during image collection to ensure that local contexts were accurately captured. All images were newly captured specifically for this initiative. Image collection was completed prior to the speech data collection, and all images underwent rigorous quality control procedures, including automated and manual checks, to ensure high quality and the absence of any personally identifiable information (PII). During the data collection process, speakers were shown up to 50 images and asked to verbally describe whatever came to mind upon viewing each image in the language spoken at home. Data collection was conducted through a mobile application. The collected data underwent a multi-stage quality control process as shown in Figure \ref{fig:pipeline}. Initially, audio recordings were evaluated for quality and format consistency. Data collection was carried out with the support of vendors who coordinated the process at the district level and were responsible for the first level of quality control.

The manually generated transcriptions for each audio segment undergo consistency checks and automated quality validation. Each transcription is subsequently reviewed independently by separate transcribers for correction, quality assurance, and verification of annotated audio events. Throughout the process, well-defined transcription guidelines are strictly followed. Transcribers for each district are selected from the same district to ensure familiarity with local dialects and linguistic variations. Code-switched content is transcribed using both the native script and the original script, while non-speech audio events and noise instances are explicitly annotated alongside the spoken content. Following the initial correction stage, transcription supervisors randomly audit approximately 10\% of the transcriptions. Identified issues are returned to the respective transcribers for correction. If systematic or recurring errors are observed in a transcriber's work, the complete set of data assigned to that transcriber is reassigned to another transcriber for comprehensive review and correction. The corrected transcriptions subsequently undergo another round of automated validation, followed by a consistency analysis across the three independent transcriptions. If substantial discrepancies are detected among the three transcriptions for an audio segment, the segment is sent for manual adjudication by three independent transcribers.

Following rigorous quality checks, the Vaani benchmark dataset comprises 5,050 speech utterances from 1,062 unique speakers, totaling 10.91 hours of audio collected across 102 districts in 20 states and 2 Union Territories.\footnote{\url{https://huggingface.co/datasets/ARTPARK-IISc/Vaani-Benchmark-V1.0}} Utterances have an average duration of 7.8~s (median: 7.2~s; range: 1.3--20.0~s) and are uniformly encoded as 16~kHz, mono, 16-bit PCM audio. The benchmark references 3,701 unique images. The three reference transcriptions are byte-identical for only 0.6\% of utterances and remain distinct for 75.5\% even after normalizing annotation markup and punctuation, motivating the multi-reference evaluation protocol adopted in this work. The benchmark is approximately balanced by gender at both the speaker level (536 male, 526 female) and utterance level (2,528 male, 2,522 female). The data span diverse demographic, linguistic, geographic, and recording conditions, including variation in age (20--70 years; median: 32), education, socio-economic stratum, multilingualism (42 languages), recording device, years of residence, and geographic origin. Speakers are distributed across 567 distinct PIN codes spanning multiple states and districts. The benchmark is openly licensed under CC BY 4.0.

\section{Evaluations}
\label{sec:eval}

We evaluated multiple models on the proposed benchmark dataset, including both open-source and commercial systems. We evaluate two tasks: automatic speech recognition, reported as WER under three evaluation protocols, and image retrieval, in which a spoken (or transcribed) description is used to retrieve the image that elicited it.

\begin{table}[!t]
\centering
\caption{WER Scores for Different ASR Models}
\label{tab:wer_scores}
\setlength{\tabcolsep}{4pt}
\begin{tabular}{cllccc}
\hline
\textbf{S.No} & \textbf{Model} & \textbf{Type} &
\multicolumn{3}{c}{\textbf{WER (\%)}} \\
\cline{4-6}
& & &
\shortstack{\textbf{App.}\\\textbf{1}} &
\shortstack{\textbf{App.}\\\textbf{2}} &
\shortstack{\textbf{App.}\\\textbf{3}} \\
\hline
1 & Gemini-3.1-Pro\cite{comanici2025gemini} & Closed & 19.0 & 15.6 & 12.3 \\
2 & Sarvam Saaras v3\cite{sarvam_asr_2026} & Closed & 19.5 & 16.4 & 13.1 \\
3 & \shortstack[l]{Indic-conformer-600m-\\multilingual\cite{ai4bharat_indicconformer600m2025}} & Open & 20.0 & 16.8 & 13.1 \\
4 & Gemini-3.5-Flash\cite{google_gemini3flash_2025} & Closed & 22.0 & 18.7 & 15.3 \\
5 & Google Chirp 3\cite{google_chirp_speech2025} & Closed & 22.5 & 19.2 & 15.6 \\
6 & Shrutam-HindiASR-1.0\cite{bharatgen2025shrutam} & Open & 24.5 & 21.5 & 18.1 \\
7 & Gemma4-E2B-IT\cite{team2026gemma} & Open & 26.2 & 23.2 & 19.4 \\
8 & Azure Speech\cite{microsoft_azure_speech2025} & Closed & 26.6 & 23.6 & 18.7 \\
9 & OmniASR\_LLM\_1B\cite{omnilingualasrteam2025omnilingualasropensourcemultilingual} & Open & 32.2 & 29.3 & 25.6 \\
10 & Whisper-large-v3\cite{radford2022whisper} & Open & 32.8 & 30.0 & 26.4 \\
\hline
\end{tabular}
\end{table}

\subsection{Automatic Speech Recognition}
\label{ssec:asr}

\begin{sloppypar}
We evaluate open-source and commercial systems spanning Indic-focused
encoder--decoder models \cite{ai4bharat_indicconformer600m2025}; large multilingual
ASR models \cite{radford2022whisper,omnilingualasrteam2025omnilingualasropensourcemultilingual};
multimodal LLM-based systems \cite{google_gemini3flash_2025, team2026gemma, comanici2025gemini}; and commercial
cloud APIs \cite{google_chirp_speech2025,microsoft_azure_speech2025,sarvam_asr_2026}.
Table~\ref{tab:wer_scores} lists the full set.
\end{sloppypar}

We evaluate ASR systems under three approaches. In Approach~1, corpus
WER is computed separately against each transcription set (Set-1,
Set-2, Set-3), and the minimum is reported. In Approach~2, the minimum
error count over the three references is selected for each segment,
and these counts are aggregated over the corpus, normalised by the length of the selected reference. Approach~3 models
word-level variation across annotators: the hypothesis is aligned
independently to each reference by minimum edit distance. A hypothesis
word is correct if it matches in any alignment, a substitution if it
is substituted in at least one, and an insertion only if it is
inserted in all three; a reference word is a deletion only if deleted
in all three. The denominator is $N = C + S + D$, the length of the
implied (materialised) reference, with errors aggregated over the
corpus. An error is thus counted only when the output matches no
acceptable realisation in any reference. The three transcription sets
exhibit pairwise WERs of 10.40--13.34\%; scoring each set against the
other two under Approach~3 yields 6.63\%, quantifying inter-annotator
variation and providing context for ASR WERs.
\begin{figure}[!t]
\centering
\includegraphics[width=\linewidth]{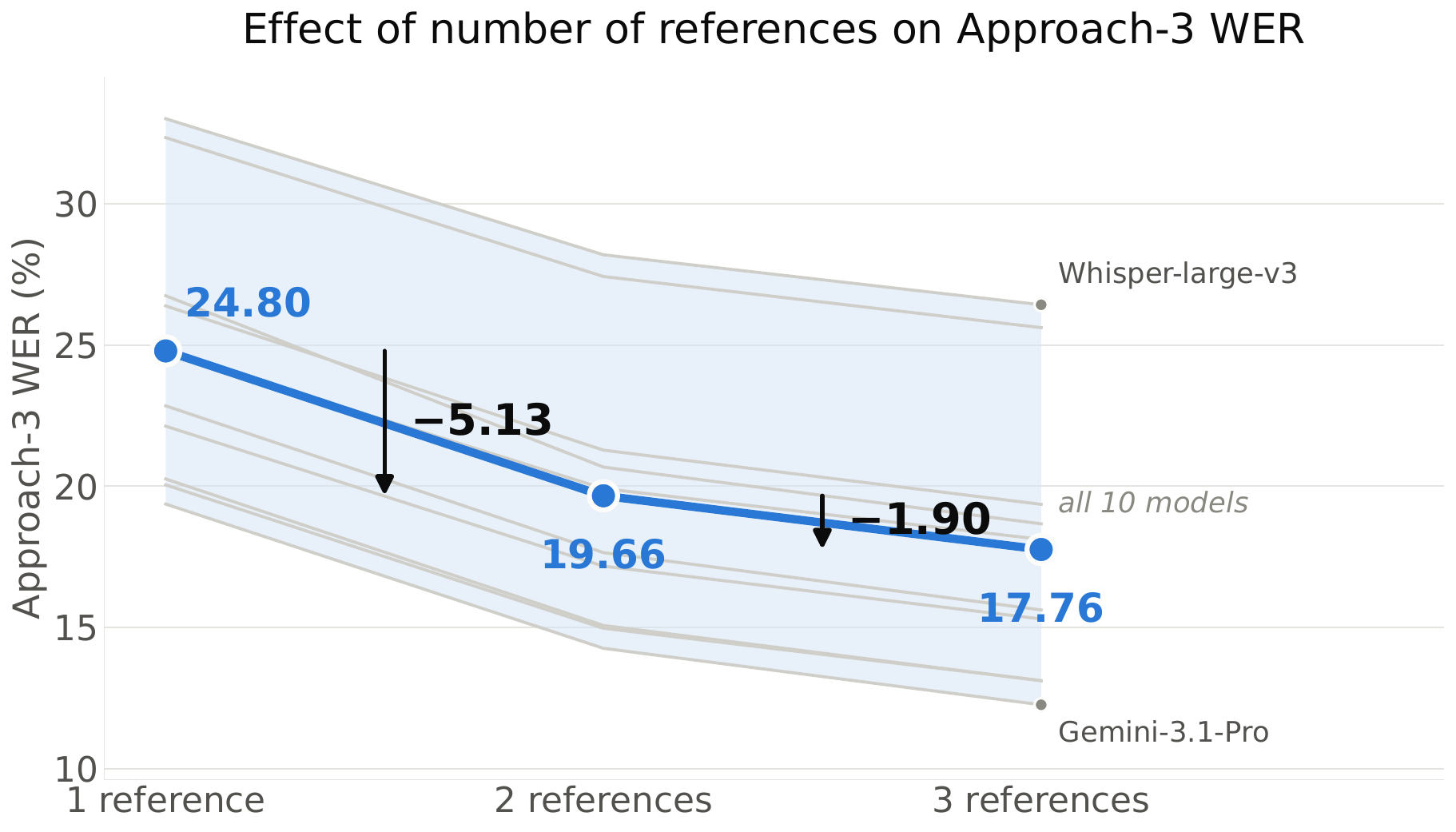}
\caption{WER variations against \#References}
\label{fig:ref_count}
\end{figure}

After evaluation, the WERs obtained using Approaches 1, 2, and 3 are presented in Table~\ref{tab:wer_scores}. The differences across the three approaches highlight the effect of multiple valid transcriptions for the same audio segment. Figure~\ref{fig:ref_count} further examines the benefit of adding reference transcriptions under Approach 3. Each system is evaluated using all possible reference subsets of a given size, and the results are averaged to avoid dependence on a particular transcription set. Across the ten systems, the average WER decreases from 24.80\% with one reference to 19.66\% with two references and 17.76\% with three references. Thus, the second reference provides a larger reduction of 5.13 percentage points, while the third provides a smaller reduction of 1.90 points. The reduction from the second to the third reference is also consistent across systems, ranging from 1.77 to 2.03 points. System rankings remain unchanged as references are added, and the gap between the best and worst systems remains similar. These results show diminishing benefits from additional references and support the use of three references as a reasonable choice for the benchmark.

\begin{figure}[t]
\centering
\includegraphics[width=\linewidth]{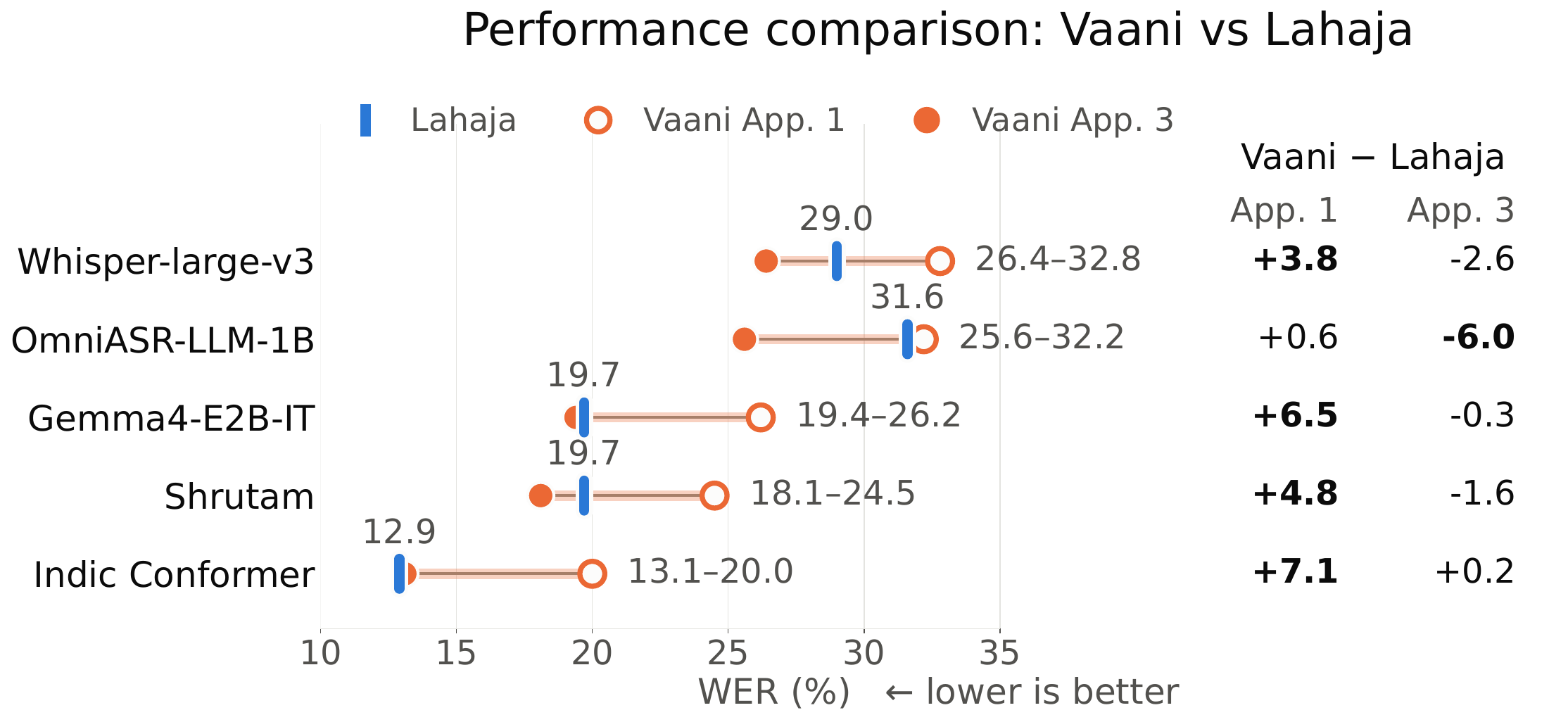}
\caption{WER comparison: Vaani vs Lahaja}
\label{fig:lahaja_vs_vaani}
\end{figure}

To verify that Approach~3 does not credit words absent from a valid
transcription, we materialise the per-utterance reference implicitly
selected for the Gemini-3.1-Pro hypotheses. Computing standard
single-reference WER against this materialised reference reproduces
the Approach-3 score exactly over the full corpus, confirming that it
is the reference Approach~3 scores against. We then manually validated
a uniform random 30-minute sample of 242 utterances (1,811\,s, 214
speakers, 81 districts across 20 states and 2 UTs): annotators listened
to each recording and verified that the materialised reference was a
correct transcription of the audio. Of these, 241 (99.6\%) were
confirmed correct with no corrections. The single flagged case arose
from an empty ASR hypothesis, where deleted reference words cannot be
recovered in order; this affects only 2 of 5,050 utterances and does
not affect the count-based WER.

\begin{figure}[t]
\centering
\includegraphics[width=\linewidth]{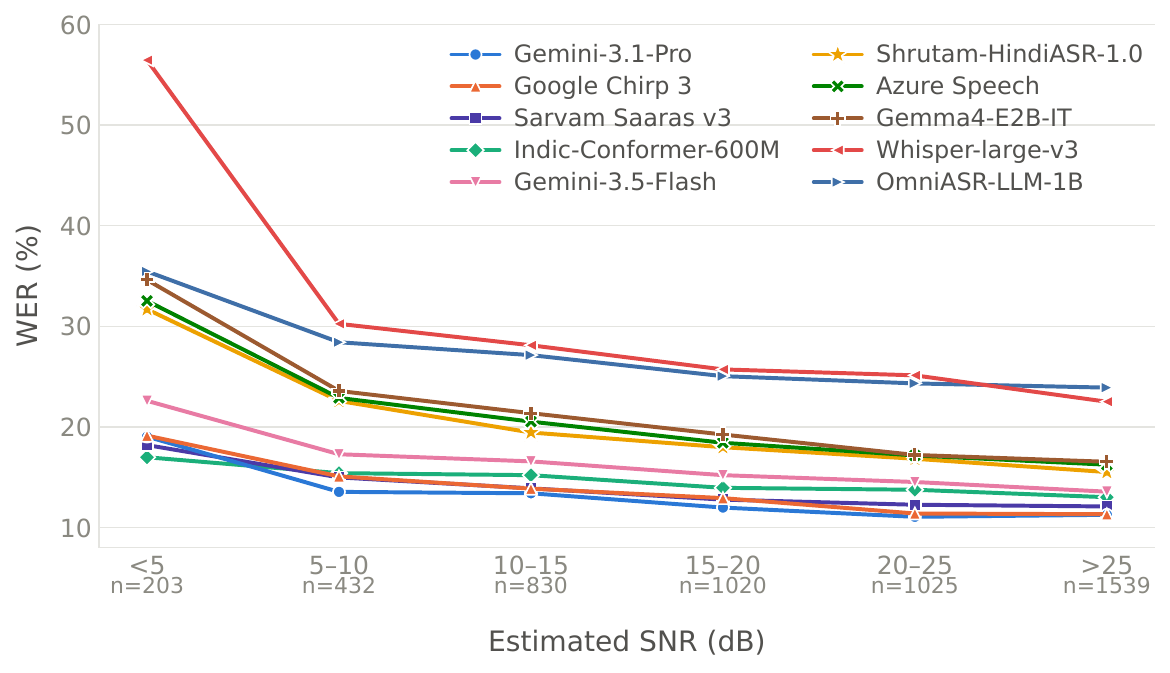}
\caption{WER variations against the estimated SNR}
\label{fig:fig_wer_vs_snr}
\end{figure}

\begin{figure}[t]
\centering
\includegraphics[width=\linewidth]{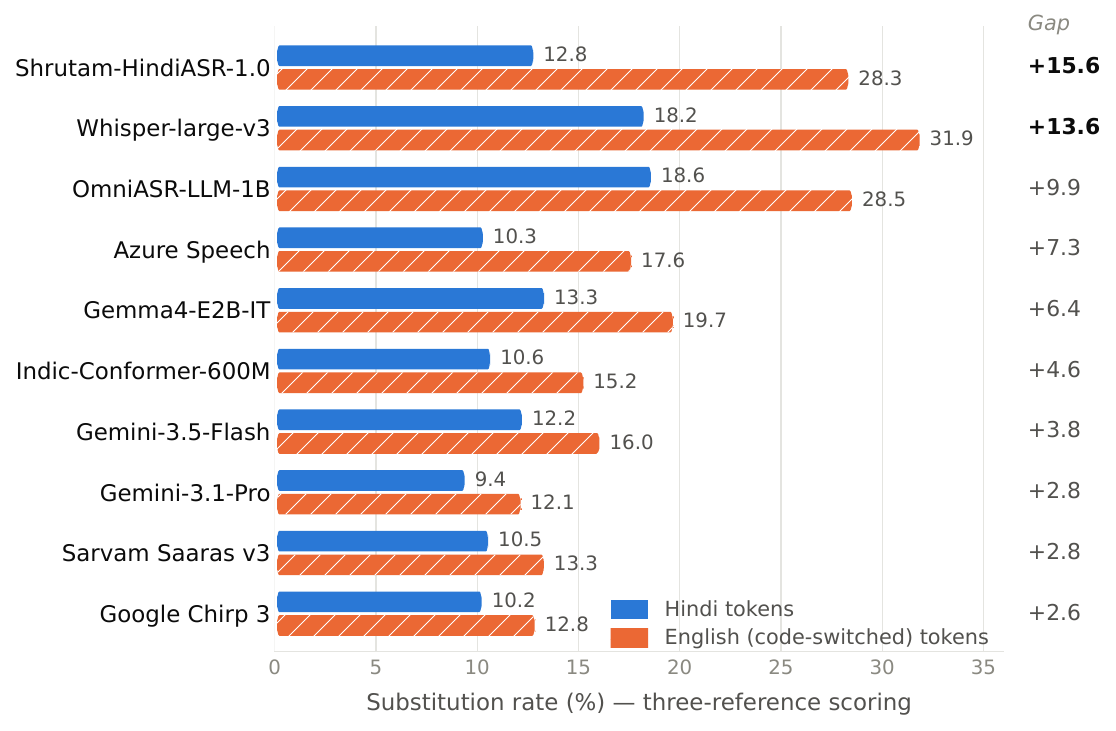}
\caption{Errors in Hindi tokens vs English (code-switched) tokens}
\label{fig:hindi_vs_english}
\end{figure}

Figure~\ref{fig:lahaja_vs_vaani} compares five systems on Lahaja and Vaani using identical
normalisation; Lahaja uses standard corpus WER over 6,152 utterances.
Under Approach~1, Vaani is harder than Lahaja for all five systems
(+0.6 to +7.1 points), with IndicConformer showing the largest gap.
Under Approach~3, the gap narrows or reverses (+0.2 to $-$6.0 points),
indicating that part of the apparent difficulty reflects valid
transcription variation rather than recognition errors. Model
rankings also differ across benchmarks, highlighting limited
generalisation from a single Hindi benchmark.
The datasets provide complementary geographic coverage, sharing only 12 districts and covering 173 distinct districts combined. We also evaluate performance across estimated SNR bands (Figure~\ref{fig:fig_wer_vs_snr}). Since no clean reference exists, per-utterance SNR is estimated using the pretrained DNS64 denoiser \cite{defossez2020real}, treating the residual as noise. WER decreases with increasing SNR, but degradation varies substantially: from $>25$ dB to $<5$ dB, IndicConformer degrades by only 3.7 points, compared with 34.1 points for Whisper-large-v3. Vaani references mark code-switched English words with their source forms, enabling separate analysis of English and Hindi tokens. We report substitution rates only, scored against all three references using Approach 3. ASR hypotheses are evaluated without script conversion, using the corresponding Latin- or Devanagari-script representation in the reference for matching. All systems perform worse on code-switched tokens (Figure~\ref{fig:hindi_vs_english}), but the penalty varies substantially: Shrutam-HindiASR-1.0 degrades by 15.6 points (12.8\% to 28.3\%), compared with 2.6 points for Google Chirp 3 (10.2\% to 12.8\%).

\begin{figure}[t]
\centering
\includegraphics[width=\linewidth]{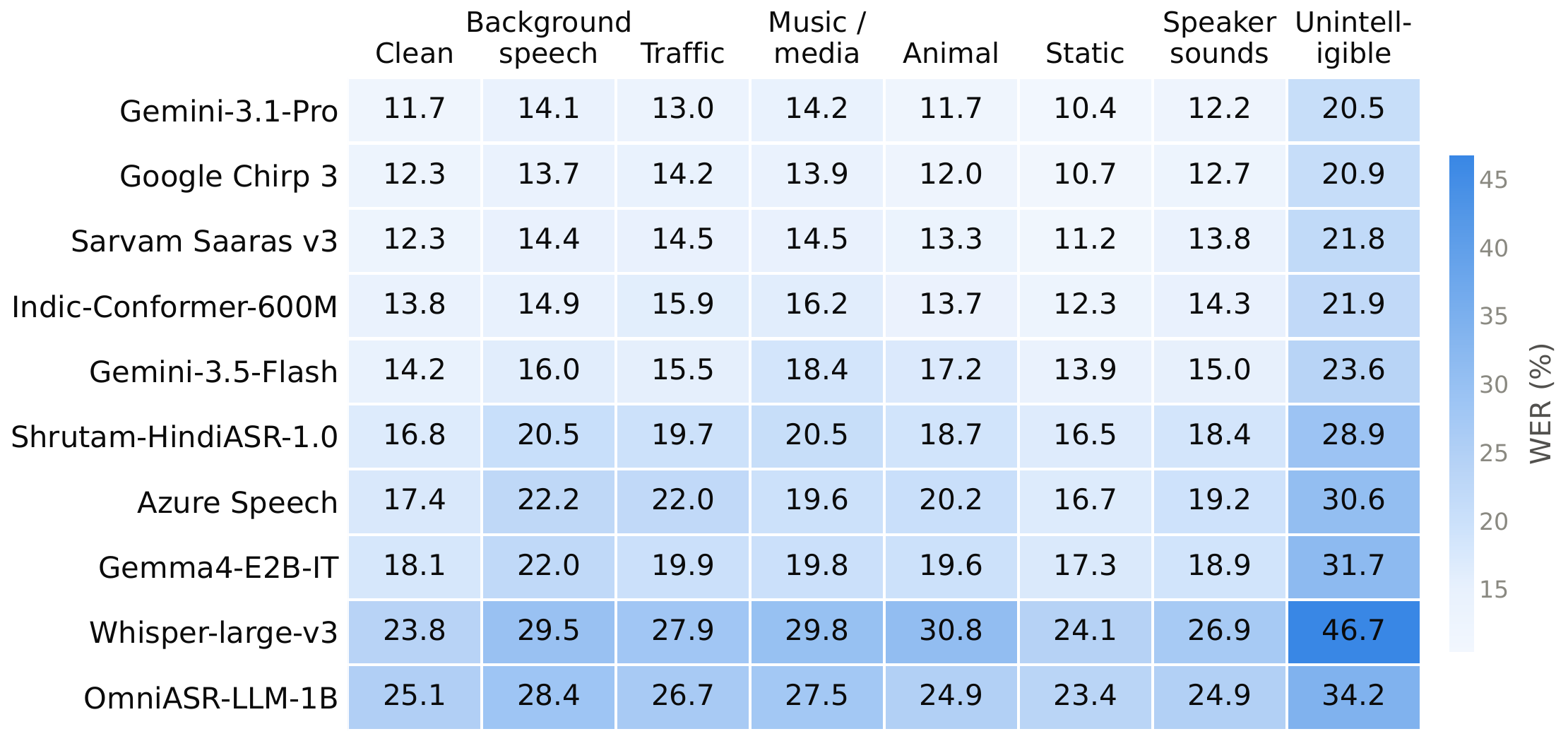}
\caption{WER against various types of noise present}
\label{fig:fig_noise_matrix}
\end{figure}

Figure~\ref{fig:fig_noise_matrix} breaks Approach-3 WER down by acoustic condition. Categories are based on annotators’ inline event tags grouped by keyword; tags are multi-label, averaging 1.61 per utterance, so categories overlap, while Clean denotes the 1,857 utterances without environmental tags. Since the categories contain different utterances, WER differences should be interpreted through within-category model comparisons rather than as direct estimates of noise penalties. Background speech is particularly challenging, with Chirp 3 performing best and Whisper-large-v3 showing substantially higher WER. Static/electrical noise yields relatively low WER for most systems, likely reflecting the nature of the tagged recordings rather than an inherently easier condition. Unintelligible is not a noise type but a difficulty marker for speech that annotators could not resolve, and is retained as a reference for challenging speech.


\subsection{Image Retrieval}
\label{ssec:retrieval}

We evaluate speech-to-image and text-to-image retrieval on the full test set of $5{,}050$ utterances against $3{,}701$ images. Each utterance is associated with the image that elicited it, forming image--speech--text triplets without additional annotation. We report Recall@$K$. For text-to-image retrieval, all models use the same consensus transcription, obtained as the medoid of the three human references under symmetric WER, with annotation markers removed. This ensures that differences in retrieval performance reflect differences in model representations rather than differences in textual input. We evaluate zero-shot speech--image models, as well as multilingual text--image models using the consensus transcriptions.
The best zero-shot speech model, multilingual M-SpeechCLIP\cite{berry2023m}, trained jointly on English, Hindi, and Japanese, reaches $3.33\%$ R@1, substantially outperforming the Hindi-only variant ($1.21\%$). Text-based retrieval is considerably stronger, although the task remains challenging: the best multilingual text--image model on Hindi consensus transcription, NLLB-CLIP-Large\cite{visheratin2023nllb}, reaches $13.60\%$ R@1. However, these results involve different models and modalities and therefore do not by themselves establish the source of the performance gap.

\begin{table}[!t]
\centering
\footnotesize
\setlength{\tabcolsep}{3pt}
\renewcommand{\arraystretch}{0.9}
\caption{Image retrieval on Vaani-Bench (R@K in \%; chance R@1
= 0.03). Text rows query the Hindi consensus transcription unless
marked MT En (its machine translation). Speech rows use original
recordings unless marked TTS (Kokoro speech of the Hi or MT En
text). Sp.+Text rows query with both the speech and its
corresponding text. ``tri'': En+Hi+Ja M-SpeechCLIP.}
\label{tab:image_retrieval}
\begin{tabular}{@{}llrrr@{}}
\hline
\textbf{Model} & \textbf{Query} & \textbf{R@1} & \textbf{R@5} & \textbf{R@10}\\
\hline
nllb-clip-large\cite{visheratin2023nllb} (MT En) & Text & \textbf{15.11} & \textbf{29.17} & \textbf{36.59}\\
nllb-clip-large & Text & 13.60 & 27.58 & 34.81\\
jina-clip-v2\cite{koukounas2024jina} (MT En) & Text & 13.45 & 26.48 & 34.02\\
siglip2-large\cite{tschannen2025siglip} (MT En) & Text & 10.77 & 22.36 & 29.84\\
jina-clip-v2 & Text & 10.34 & 22.48 & 29.86\\
jina-omni\cite{honicke2026jina} (MT En) & Text & 9.25 & 19.73 & 26.10\\
siglip2-base (MT En) & Text & 8.79 & 20.28 & 27.43\\
jina-omni & Text & 7.25 & 16.53 & 22.04\\
siglip2-large & Text & 5.05 & 12.42 & 17.88\\
siglip2-base & Text & 3.35 & 8.57 & 12.67\\
\hline
M-SpeechCLIP-tri\cite{berry2023m} + TTS En & Speech & 3.88 & 11.84 & 17.09\\
M-SpeechCLIP-tri + TTS Hi & Speech & 3.43 & 10.55 & 15.70\\
M-SpeechCLIP-tri & Speech & 3.33 & 9.05 & 13.92\\
jina-omni + TTS Hi & Speech & 1.70 & 4.40 & 6.32\\
jina-omni + TTS En & Speech & 1.64 & 3.82 & 5.62\\
jina-omni & Speech & 1.49 & 3.88 & 5.62\\
M-SpeechCLIP-Hi & Speech & 1.21 & 3.94 & 6.81\\
\hline
jina-omni + TTS En & Sp.+Text & 7.61 & 17.23 & 22.36\\
jina-omni + TTS Hi & Sp.+Text & 6.02 & 13.43 & 18.38\\
jina-omni & Sp.+Text & 5.66 & 13.35 & 18.30\\
\hline
\end{tabular}
\end{table}

 To probe whether recording acoustics contribute to the low speech-retrieval performance, we re-synthesise the test utterances as clean Hindi and English speech using Kokoro TTS\cite{hexgrad2025kokoro}, and evaluate them with jina-embeddings-v5-omni-nano\cite{honicke2026jina}.
 English queries are machine translations of the same human transcriptions, produced with gpt-5.6-luna \cite{openai2026gpt56luna}.
Speech-to-image performance remains similar across the three conditions:
$1.49\%$ R@1 for the original recordings, $1.70\%$ for synthetic Hindi speech,
and $1.64\%$ for synthetic English speech, with no pair significantly different
under a paired exact McNemar test on top-1 outcomes ($p=0.28$ real vs.\
synthetic Hindi, $0.53$ real vs.\ synthetic English, $0.85$ synthetic Hindi
vs.\ synthetic English). In contrast, text-to-image retrieval improves from
$7.25\%$ for Hindi text to $9.25\%$ for English text, with similar trends
across the evaluated text encoders. These results suggest that the acoustic characteristics of the recordings alone do not explain the low speech-retrieval performance, pointing instead to limitations in the current speech representations in capturing image-relevant semantic information.

\section{Conclusion}
\label{sec:conclusion}
We present an inclusive, multimodal Hindi ASR benchmark comprising speech data from 102 districts, with three independent reference transcriptions for each segment. Evaluation of open-source and proprietary ASR systems shows that multi-reference scoring can substantially change WER compared with single-reference evaluation, highlighting the importance of accounting for valid transcription variation. Analysis of noise, SNR, and code-switched speech further reveals system-level differences that are not captured by aggregate WER alone. The image retrieval experiments show that text queries consistently outperform speech queries from the same models, while English translations of Hindi text achieve higher retrieval performance than the original Hindi. These findings highlight the need for robust evaluation protocols and better speech and Hindi-language representations for multimodal speech understanding.

\bibliographystyle{IEEEbib}
\bibliography{mybib}

\end{document}